\documentclass[a4paper,10pt]{article}
\usepackage{stmaryrd}
\usepackage{amsfonts}
\usepackage{bbm}
\usepackage{amscd}
\usepackage{mathrsfs}
\usepackage{latexsym,amssymb,amsmath,amscd,amscd,amsthm,amsxtra,xypic}
\usepackage[dvips]{graphicx}
\usepackage[utf8]{inputenc}
\usepackage[T1]{fontenc}
\usepackage{lmodern}
\usepackage{amssymb}
\usepackage[all]{xy}
\usepackage{nicefrac,mathtools,enumitem}
\usepackage{microtype}
\usepackage{lmodern}

\newcommand {\emptycomment}[1]{}
\newtheorem{thm}{Theorem}[section]

\newtheorem{pro}[thm]{Proposition}
\newtheorem{ex}[thm]{Example}
\newtheorem{rmk}[thm]{Remark}
\newtheorem{defi}[thm]{Definition}

\newcommand{\lon }{\,\rightarrow\,}
\newcommand{\be }{\begin{equation}}
\newcommand{\ee }{\end{equation}}

\newcommand{\pf}{\noindent{\bf Proof.}\ }

\newcommand{\g}{\frkg}

\newcommand{\frka}{\mathfrak a}
\newcommand{\frkb}{\mathfrak b}

\newcommand{\frkg}{\mathfrak g}
\newcommand{\frkh}{\mathfrak h}

\newcommand{\frkl}{\mathfrak l}

\def\qed{\hfill ~\vrule height6pt width6pt depth0pt}

\newcommand{\br}[1]{   [ \cdot,    \cdot  ]   }

\newcommand{\Ad}{\mathrm{Ad}}

\newcommand{\gl}{\mathfrak {gl}}

\newcommand{\ad}{\mathrm{ad}}

\begin{document}
\title{
{Skew-Hom-Lie algebras in Semi-Euclidean spaces
} }
\author{Zhen Xiong \\
Department of Mathematics, Yichun University,
 Yichun,Jiangxi 336000,  China
\\\vspace{3mm}
email: 205137@jxycu.edu.cn }
\date{}
\footnotetext{Supported by the Science and Technology Project(GJJ190832)of Department of Education, Jiangxi Province.}
\maketitle

\begin{abstract}
  In this paper, first we introduce the notion of a skew-Hom-Lie algebra and give some examples. Then we study their representations and give the coboundary operator of skew-Hom-Lie algebras. As an application, there have a skew-Hom-Lie algebra $(R^4_2,[\cdot,\cdot]_\theta,P)$ in semi-Euclidean spaces. For the null space of Semi-Euclidean spaces, there have a subset $V^*$ of the null space, and  $V^*$ is invariant under the actions of $[\cdot,\cdot]_\theta$ and $P$.

\end{abstract}


\section{Introduction}

The notion of Hom-Lie algebras was introduced by Hartwig, Larsson,
and Silvestrov in \cite{HLS} as part of a study of deformations of
the Witt and the Virasoro algebras. In a Hom-Lie algebra, the Jacobi
identity is twisted by a linear map, called the Hom-Jacobi identity.
Some $q$-deformations of the Witt and the Virasoro algebras have the
structure of a Hom-Lie algebra \cite{HLS,hu}. Because of close relation
to discrete and deformed vector fields and differential calculus
\cite{HLS,LD1,LD2}, more people pay special attention to this algebraic structure\cite{AEM,homlie1,Ca,hom-Lie algebroids,zh}. If $(\frka,[\cdot,\cdot],\gamma)$ is a Hom-Lie algebra, $k$ is a real number,when $k\neq 1$, then $(\frka,[\cdot,\cdot],k\gamma)$ is not a Hom-Lie algebra. On the other hands,  the Semi-Euclidean space is a vector space with pseudoscalar product which is different from Euclidean space. The study of semi-Euclidean spaces has produced fruitful results \cite{AE,EG,H,mk,jb}. It is well known that there exist spacelike submanifolds, timelike submanifolds, and null submanifolds in semi-Euclidean space. Null submanifolds appear in many physics papers. For example, the null submanifolds are of interest because they provide models of different horizon types such as event horizons of Kerr black holes, Cauchy horizons, isolated horizons, Kruskal horizons, and Killing horizons\cite{G1,ej,sw,mj,vj,kp}. Null submanifolds are also studied in the theory of electromagnetism.

In this paper, first we find out a kind of deformations of Lie algebras, these deformations are associated with Hom-Lie algebras. Then, we give some properties of these deformations and study their representations. Next, we give the coboundary operator of  these deformations. Last. we find out these deformations in null spaces of Semi-Euclidean spaces.

The paper is organized as follows. In Section 2, we introduce the notion of skew-Hom-Lie algebras, study their representations and give some properties of skew-Hom-Lie algebras. In Section 3,  we give the coboundary operator of  skew-Hom-Lie algebras. In Section 4, we construct a series of skew-Hom-Lie algebras $(R^4_2,[\cdot,\cdot]_\theta,P)$ in Semi-Euclidean 4-spaces, then, we prove that for the null space of Semi-Euclidean 4-spaces , there exist a subset $V^*$ of the null space, and $V^*$ is invariant under the actions of $[\cdot,\cdot]_\theta$ and $P$.

\section{Skew-Hom-Lie algebras and their representations}

The notion of a Hom-Lie algebra was introduced in \cite{HLS}, see also \cite{BM,MS2} for more information.

\begin{defi}
\begin{itemize}
\item[\rm(1)]
  A Hom-Lie algebra is a triple $(\frka,\br ,,\gamma)$ consisting of a
  vector space $\frka$, a skew symmetric bilinear map (bracket) $\br,:\wedge^2\frka\longrightarrow
  \frka$ and a linear transformation $\gamma:\frka\lon\frka$ satisfying $\gamma[x,y]=[\gamma(x),\gamma(y)]$, and the following Hom-Jacobi
  identity:
  \begin{equation*}
   [[y,z],\gamma(x)]+[[z,x],\gamma(y)]+[[x,y],\gamma(z)]=0,\quad\forall
x,y,z\in\frkg.
  \end{equation*}

 A Hom-Lie algebra is called a regular Hom-Lie algebra if $\gamma$ is
a linear automorphism.
\item[\rm(2)] A subspace $\frkb\subset\frka$ is a Hom-Lie sub-algebra of $(\frka,\br ,,\gamma)$ if
 $\gamma(\frkb)\subset\frkb$ and
  $\frkb$ is closed under the bracket operation $\br,$, i.e. for all $ x,y\in\frkb$,
  $[x,y] \in\frkb.  $

   \item[\rm(3)] A morphism from the  Hom-Lie algebra$(\frka,[\cdot,\cdot]_{\frka},\gamma)$
 to the Hom-Lie algebra$(\frkb,[\cdot,\cdot]_{\frkb},\delta)$
 is a linear map
$\psi:\frka\longrightarrow\frkb$ such that
$\psi([x,y]_{\frka})=[\psi(x),\psi(y)]_{\frkb}$ and
$\psi\circ \gamma =\delta\circ \psi$.
\end{itemize}
\end{defi}
When $(\frka,\br ,,\gamma)$ is a Hom-Lie algebra, then $\gamma^k([x,y])=[\gamma^k(x),\gamma^k(y)]$, $k=1,2,3,\cdots$. Now, we give the notion of skew-Hom-Lie algebras.
\begin{defi}
\begin{itemize}
\item[\rm(1]
  A skew-Hom-Lie algebra is a triple $(\frkg,\br ,,\beta)$ consisting of a
  vector space $\frkg$, a skew symmetric bilinear map (bracket) $\br,:\wedge^2\frkg\longrightarrow
  \frkg$ and a linear transformation $\beta:\frkg\lon \frkg$ satisfying $\beta([x,y])=-[\beta(x),\beta(y)]$, and the following Hom-Jacobi
  identity:
  \begin{equation}
   [[y,z],\beta(x)]+[[z,x],\beta(y)]+[[x,y],\beta(z)]=0,\quad\forall
x,y,z\in V.
  \end{equation}
A skew-Hom-Lie algebra is called a regular skew-Hom-Lie algebra if $\beta$ is
a linear automorphism.
\item[\rm(2] A subspace $\frkh\subset\frkg$ is a skew-Hom-Lie sub-algebra of $(\frkg,\br ,,\beta)$ if
 $\beta(\frkh)\subset\frkh$ and
  $\frkh$ is closed under the bracket operation $\br,$, i.e. for all $ x,y\in\frkh$,
  $[x,y] \in\frkh.  $
\item[\rm(3] A morphism from the  skew-Hom-Lie algebra
$(\frkg,[\cdot,\cdot]_{\frkg},\beta)$ to the skew-Hom-Lie algebra
$(\frkh,[\cdot,\cdot]_{\frkh},\lambda)$ is a linear map
$f:\frkg\longrightarrow\frkh$ such that
$f([x,y]_{\frkg})=-[f(x),f(y)]_{\frkh}$ and
$f\circ \beta =\lambda\circ f$.
  \end{itemize}
\end{defi}
\begin{rmk}
\begin{itemize}
\item[\rm 1)]
Let $(\frka,[\cdot,\cdot],\gamma)$ be a Hom-Lie algebra,when $\gamma$ is an identity,then $(\frka,[\cdot,\cdot],\gamma)$ is a Lie algebra; but when
$(\frkg,\br ,,\beta)$ is a skew-Hom-Lie algebra, map $\beta$ can not be an identity; so a Lie algebra is a Hom-Lie algebra, but it is not a skew-Hom-Lie algebra;
\item[\rm 2)]
$(\frka,[\cdot,\cdot],\gamma)$ is a Hom-Lie algebra,then $(\frka,[\cdot,\cdot],(-1)\gamma)$ is a skew-Hom-Lie algebra;
\item[\rm 3)]
when $(\frkg,\br ,,\beta)$ is a skew-Hom-Lie algebra, we have: $\beta^{2k}([x,y])=[\beta^{2k}(x),\beta^{2k}(y)]$, and $\beta^{2k-1}([x,y])=-[\beta^{2k-1}(x),\beta^{2k-1}(y)]$, $k=1,2,3,\cdots.$
\end{itemize}
\end{rmk}
\begin{ex}
Let $R^3$ be a 3-dimensional vector space. For $x,y\in R^3$, $x=(x_1,x_2,x_3)^T, ,y=(y_1,y_2,y_3)^T$,we define a skew symmetric bilinear map (bracket)
\begin{equation*}[x,y]= x\wedge y=\begin{vmatrix}
e_1 & e_2 & e_3\\
x_1 & x_2 & x_3\\
y_1 & y_2 & y_3
\end{vmatrix},
\end{equation*}
where $\{e_1,e_2,e_3\}$ is the canonical basis of $R^3$. We also have the scalar product $<x,y>_1=x^Ty$. Let $A$ be a $3\times 3$-matrix, $A$ act on $R^3$by $A(x)=Ax$, then, $A$ is a linear map. When $AA^T=\rm{id}$, for $x,y,z\in R^3$, we have:
\begin{equation*}<A([x,y]),z>_1=<x\wedge y,A^Tz>_1=\begin{vmatrix}
x^T\\
y^T\\
(A^Tz)^T
\end{vmatrix}
\end{equation*}
On the other hands, we have:
\begin{equation*}
<[Ax,Ay],z>_1=<Ax\wedge Ay,z>_1=\begin{vmatrix}

(Ax)^T\\
(Ay)^T\\
z^T\end{vmatrix}
=|A|\begin{vmatrix}
x^T\\
y^T\\
(A^Tz)^T
\end{vmatrix}
=|A|<A([x,y]),z>_1.
\end{equation*}
So, there have:$$|A|=1, \quad A([x,y])=[A(x),A(y)];$$
 $$|A|=-1,\quad A([x,y])=-[A(x),A(y)].$$
 Let $$[x,y]_1=A([x,y]),$$we have:
\begin{itemize}
\item[\rm(1)]
when $|A|=1$, $(R^3,[\cdot,\cdot]_1,A)$ is a Hom-Lie algebra;
\item[\rm(2)]
when $|A|=-1$, $(R^3,[\cdot,\cdot]_1,A)$ is a skew-Hom-Lie algebra.
\end{itemize}
\end{ex}
\begin{pro}
Let $\alpha\in \frkg\frkl(V)$ and $\alpha^2=-\rm{id}$, we define a linear map $$Ad_\alpha:\frkg\frkl(V)\longrightarrow\frkg\frkl(V)$$ by $Ad_\alpha(B)=\alpha B\alpha$, and
a bilinear map (bracket)$$[A,B]_\alpha=\alpha A\alpha B\alpha-\alpha B\alpha A\alpha,$$ then $(\frkg\frkl(V),[\cdot,\cdot]_\alpha,Ad_\alpha)$ is a skew-Hom-Lie algebra.
\end{pro}
\pf Obviously,$[\cdot,\cdot]$ is a skew bilinear map.  More, we have:
\begin{eqnarray*}
Ad_\alpha([A,B]_\alpha) &=&\alpha^2 A\alpha B\alpha^2-\alpha^2
B\alpha A\alpha^2\\
&=& A\alpha B-B\alpha A.
\end{eqnarray*}
On the other hands,
\begin{eqnarray*}
[Ad_\alpha(A),Ad_\alpha(B)]_\alpha &=&\alpha^2 A\alpha^3 B\alpha^2-\alpha^2
B\alpha^3 A\alpha^2\\
&=& -A\alpha B+B\alpha A.
\end{eqnarray*}
So, $Ad_\alpha([A,B]_\alpha)=-[Ad_\alpha(A),Ad_\alpha(B)]_\alpha$.\\
For all $A,B,C\in\frkg\frkl(V)$, we have
\begin{eqnarray*}
&&[[A,B]_\alpha,\Ad_\alpha(C)]_\alpha+\circlearrowleft_{A,B,C}\\
&=&[\alpha A\alpha B\alpha,\alpha C\alpha]_\alpha-[\alpha
B\alpha A\alpha,\alpha C\alpha]_\alpha+\circlearrowleft_{A,B,C}\\
&=&-A\alpha B\alpha C+C\alpha A\alpha B+B\alpha A\alpha C-C\alpha B\alpha A+\circlearrowleft_{A,B,C}\\
&=&0.
\end{eqnarray*}
where $\circlearrowleft_{A,B,C}$ denotes summation over the cyclic permutation on $A,B,C$.
 Thus, $(\frkg\frkl(V),[\cdot,\cdot]_\alpha,Ad_\alpha)$ is a skew-Hom-Lie
algebra. \qed
\begin{rmk}
$(\frkg\frkl(V),[\cdot,\cdot]_\alpha,Ad_\alpha^2)$ is not a Hom-Lie algebra. By straightforward computations,
\begin{eqnarray*}
&&[[A,B]_\alpha,\Ad_\alpha^2(C)]_\alpha+\circlearrowleft_{A,B,C}\\
&=&[\alpha A\alpha B\alpha, C]_\alpha-[\alpha
B\alpha A\alpha, C]_\alpha+\circlearrowleft_{A,B,C}\\
&=&A\alpha BC\alpha-\alpha CA\alpha B-B\alpha AC\alpha+\alpha CB\alpha A+\circlearrowleft_{A,B,C}\\
&\neq&0.
\end{eqnarray*}
\end{rmk}

\begin{defi}
  A representation of the skew-Hom-Lie algebra $(\frkg,\br,,\beta)$ on
  a vector space $V$ with respect to $\phi\in\gl(V)$ is a linear map
  $\rho:\frkg\longrightarrow \gl(V)$, such that for all
  $x,y\in\frkg$, the following equalities are satisfied:
  \begin{eqnarray}
 \label{representation1} \rho (\beta(x))\circ \phi&=&-\phi\circ \rho (x);\\\label{representation2}
    \rho([x,y] )\circ
    \phi&=&\rho (\beta(x))\circ\rho (y)-\rho (\beta(y))\circ\rho (x).
  \end{eqnarray}
\end{defi}

\begin{thm}
Let $(\frkg,[\cdot,\cdot],\beta)$ be a skew-Hom-Lie algebra, $V$ is a
vector space, $\alpha^2=-\rm{id}$.
Then, $\rho:\frkg\longrightarrow\frkg\frkl(V)$ is a representation of $(\frkg,[\cdot,\cdot],\beta)$
on   $V$ with respect to $\alpha$ if and only if
$\rho:(\frkg,[\cdot,\cdot],\beta)\longrightarrow(\frkg\frkl(V),[\cdot,\cdot]_\alpha, Ad_\alpha)$
is a morphism of skew-Hom-Lie algebras.
\end{thm}
\pf
If $\rho :\frkg\longrightarrow\frkg\frkl(V)$ is a representation of
$(\frkg,[\cdot,\cdot],\beta)$ on  $V$ with respect to $\alpha$,  we have
\begin{eqnarray}
\label{eq:t1}\rho(\beta(x))\circ\alpha&=&-\alpha\circ\rho(x),\\
\label{eq:t2}\rho([x,y])\circ\alpha&=&\rho(\beta(x))\rho(y)-\rho(\beta(y))\rho(x).
\end{eqnarray}
By \eqref{eq:t1}, we deduce that
$$
\rho\circ\beta = Ad_\alpha\circ \rho.
$$
Furthermore, by \eqref{eq:t1} and \eqref{eq:t2}, we have
\begin{eqnarray*}
\rho([x,y])&=&-\rho(\beta(x))\circ\rho(y)\alpha+\rho(\beta(y))\circ\rho(x)\alpha\\
&=&-\alpha\rho(x)\alpha\rho(y)\alpha+\alpha\rho(y)\alpha\rho(x)\alpha\\
&=&-[\rho(x),\rho(y)]_\alpha.
\end{eqnarray*}
Thus, $\rho$ is morphism of skew-Hom-Lie algebras. The converse part is easy to be checked. The proof is completed.  \qed
\begin{defi}
Let $(\frkg,[\cdot,\cdot],\beta)$ be a skew-Hom-Lie algebra, the pseudo-adjoint representation $\ad^*:\g\longrightarrow\gl(\g)$, which is defined by $\ad^*_xy=-[x,y]$.
\end{defi}
\begin{rmk}Because of $\beta\neq \rm{id}$, a Lie algebra is not a skew-Hom-Lie algebra, for $x,y,z\in \frkg$
\begin{eqnarray*}
\ad^*_{[x,y]}\circ \beta(z)&=&-[[x,y],\beta(z)]\\
&=&[\beta(x),-[y,z]]+[\beta(y),[x,z]]\\
&=&-\ad^*_{\beta(x)}\circ\ad^*_y(z)+\ad^*_{\beta(y)}\circ\ad^*_x(z).
\end{eqnarray*}
\end{rmk}
\begin{pro}
 Let $(\frkg,[\cdot,\cdot],\beta)$ be a skew-Hom-Lie algebra and $\beta^2=-\rm{id}$. Then, the pseudo-adjoint representation $\ad^*:\g\longrightarrow\gl(\g)$, which is defined by $\ad^*_xy=-[x,y]$, is a morphism from $(\frkg,[\cdot,\cdot],\beta)$ to $(\gl(\g),[\cdot,\cdot]_\beta, Ad_\beta)$.
\end{pro}
\pf For $x,y,z\in\frkg$, we have:$Ad_\beta\circ\ad^*_x(y)=-[\beta(x),y]=\ad^*_{\beta(x)}(y)$, and
\begin{eqnarray*}
\ad^*_{[x,y]}(z)&=&[[x,y],\beta^2(z)]\\
&=&-[[y,\beta(z)],\beta(x)]-[[\beta(z),x],\beta(y)]\\
&=&[\beta(x),[y,\beta(z)]]-[\beta(y),[x,\beta(z)]].
\end{eqnarray*}
By straightforward computations,
\begin{eqnarray*}
\beta\circ\ad^*_x\circ\beta\circ\ad^*_y\circ\beta(z)&=&\\
&=&\beta\circ\ad^*_x\circ\beta(-[y,\beta(z)])\\
&=&\beta(-[x,[\beta(y),-z]])=[\beta(x),\beta([\beta(y),-z])]\\
&=&-[\beta(x),[y,\beta(z)]].
\end{eqnarray*}
So, we have:\quad $\ad^*_{[x,y]}=-[\ad^*_x,\ad^*_y]_{\beta}$. The proof is completed.  \qed

\section{Coboundary operators of skew-Hom-Lie algebras}
Let $(\frkg,[\cdot,\cdot],\beta)$ be a skew-Hom-Lie algebra, $V$ be a
vector space, $\rho:\frkg\longrightarrow\frkg\frkl(V)$ be a
representation of $(\frkg,[\cdot,\cdot],\beta)$ on the vector space
$V$ with respect to $\phi\in GL(V)$, where $\phi$ is invertible.

The set of {\bf
$k$-cochains} on $\frkg$ with values in $V$, which we denote by
$C^k(\frkg;V)$, is the set of skewsymmetric $k$-linear maps from
$\frkg\times\cdots\times\frkg$($k$-times) to $V$:
$$C^k(\frkg;V):=\{\eta:\wedge^k\frkg\longrightarrow V ~\mbox{is a
linear map}\}.$$
For $s=0,1,2,\ldots,$ define
$d^s:C^k(\frkg;V)\longrightarrow
C^{k+1}(\frkg;V)$ by
\begin{eqnarray*}
d^s\eta(x_1,\cdots,x_{k+1})&=&\sum_{i=1}^{k+1}(-1)^{i+1}\phi^{k+1+s}\rho(x_i)\phi^{-k-2-s}\eta(\beta(x_1),\cdots,\hat{x_i},\cdots,\beta(x_{k+1}))\\
&&+\sum_{i<j}(-1)^{i+j}\eta([x_i,x_j],\beta(x_1),\cdots,\widehat{x_{i,j}},\cdots,\beta(x_{k+1})),
\end{eqnarray*}
where $\phi^{-1}$ is the inverse of $\phi$, $\eta\in C^k(\frkg;V)$.
\begin{pro}
With the above notations, the map $d^s$ is a coboundary operator, i.e. $d^s\circ d^s=0$.
\end{pro}
\pf For any $\eta\in C^k(\frkg;V)$, by straightforward computations, we have
\begin{eqnarray*}
d^s\circ d^s\eta(x_1,\cdots,x_{k+2})&=&\sum_{i=1}^{k+2}(-1)^{i+1}\phi^{k+2+s}\rho(x_i)\phi^{-k-3-s}d^s\eta(\beta(x_1),\cdots,\hat{x_i},\cdots,\beta(x_{k+2}))\\
&&+\sum_{i<j}(-1)^{i+j}d^s\eta([x_i,x_j],\beta(x_1),\cdots,\widehat{x_{i,j}},\cdots,\beta(x_{k+2})).
\end{eqnarray*}
And\\
$d^s\eta(\beta(x_1),\cdots,\hat{x_i},\cdots,\beta(x_{k+2}))$
\begin{eqnarray*}
&=&\sum_{l<i}(-1)^{l+1}\phi^{k+1+s}\rho(\beta(x_l))\phi^{-k-2-s}\eta(\beta^2(x_1),\cdots,\widehat{x_{l,i}},\cdots,\beta^2(x_{k+2}))\\
&&+\sum_{l>i}(-1)^{l}\phi^{k+1+s}\rho(\beta(x_l))\phi^{-k-2-s}\eta(\beta^2(x_1),\cdots,\widehat{x_{i,l}},\cdots,\beta^2(x_{k+2}))\\
&&+\sum_{m<n<i}(-1)^{m+n}\eta(\beta([x_i,x_j]),\beta^2(x_1),\cdots,\widehat{x_{m,n,i}},\cdots,\beta^2(x_{k+2}))\\
&&+\sum_{m<i<n}(-1)^{m+n-1}\eta(\beta([x_i,x_j]),\beta^2(x_1),\cdots,\widehat{x_{m,i,n}},\cdots,\beta^2(x_{k+2}))\\
&&+\sum_{i<m<n}(-1)^{m+n}\eta(\beta([x_i,x_j]),\beta^2(x_1),\cdots,\widehat{x_{i,m,n}},\cdots,\beta^2(x_{k+2})).
\end{eqnarray*}
At the same time, we have\\
$d^s\eta([x_i,x_j],\beta(x_1),\cdots,\widehat{x_{i,j}},\cdots,\beta(x_{k+2}))$
\begin{eqnarray}
&=&\phi^{k+1+s}\rho([x_i,x_j])\phi^{-k-2-s}\eta(\beta^2(x_1),\cdots,\widehat{x_{i,j}},\cdots,\beta^2(x_{k+2}))\nonumber\\
&&+\sum_{p<i<j}(-1)^{p}\phi^{k+1+s}\rho(\beta(x_p))\phi^{-k-2-s}\eta(\beta([x_i,x_j]),\beta^2(x_1),\cdots,\widehat{x_{p,i,j}},\cdots,\beta^2(x_{k+2}))\nonumber\\
&&+\sum_{i<p<j}(-1)^{p+1}\phi^{k+1+s}\rho(\beta(x_p))\phi^{-k-2-s}\eta(\beta([x_i,x_j]),\beta^2(x_1),\cdots,\widehat{x_{i,p,j}},\cdots,\beta^2(x_{k+2}))\nonumber\\
&&+\sum_{i<j<p}(-1)^{p}\phi^{k+1+s}\rho(\beta(x_p))\phi^{-k-2-s}\eta(\beta([x_i,x_j]),\beta^2(x_1),\cdots,\widehat{x_{i,j,p}},\cdots,\beta^2(x_{k+2}))\nonumber\\
&&+\sum_{q<i<j}(-1)^{1+q}\eta([[x_i,x_j],\beta(x_q)],\beta^2(x_1),\cdots,\widehat{x_{q,i,j}},\cdots,\beta^2(x_{k+2}))\label{eq1}\\
&&+\sum_{i<q<j}(-1)^{q}\eta([[x_i,x_j],\beta(x_q)],\beta^2(x_1),\cdots,\widehat{x_{i,q,j}},\cdots,\beta^2(x_{k+2}))\label{eq2}\\
&&+\sum_{i<j<q}(-1)^{1+q}\eta([[x_i,x_j],\beta(x_q)],\beta^2(x_1),\cdots,\widehat{x_{i,j,q}},\cdots,\beta^2(x_{k+2}))\label{eq3}\\
&&+\sum_{m<n<i<j}(-1)^{m+n}\eta([\beta(x_m),\beta(x_n)],\beta([x_i,x_j]),\beta^2(x_1),\cdots,\widehat{x_{m,n,i,j}},\cdots,\beta^2(x_{k+2}))\label{eq4}\\
&&+\cdots\nonumber
\end{eqnarray}
By Hom-Jacobi  identity:
$$(\ref{eq1})+(\ref{eq2})+(\ref{eq3})=0,$$
and we have: $(\ref{eq4})+\cdots=0.$

By $\rho(\beta(x))\phi=-\phi\rho(x)$ and $\rho(\beta(x))=-\phi\rho(x)\phi^{-1}$, $\beta([x_i,x_j])=-[\beta(x_i),\beta(x_j)]$, we have\\
$d^s\circ d^s\eta(x_1,\cdots,x_{k+2})$
\begin{eqnarray}
&=&\sum_{l<i}(-1)^{l+i}\phi^{k-1+s}\rho(\beta^3(x_i))\rho(\beta^2(x_l))\phi^{-k-1-s}\eta(\beta^2(x_1),\cdots,\widehat{x_{l,i}},\cdots,\beta^2(x_{k+2}))\label{eq5}\\
&&+\sum_{l>i}(-1)^{l+i+1}\phi^{k-1+s}\rho(\beta^3(x_i))\rho(\beta^2(x_l))\phi^{-k-1-s}\eta(\beta^2(x_1),\cdots,\widehat{x_{i,l}},\cdots,\beta^2(x_{k+2}))\label{eq6}\\
&&+\sum_{m<n<i}(-1)^{m+n+i}\phi^{k+2+s}\rho(x_i)\phi^{-k-3-s}\eta([\beta(x_i),\beta(x_j)],\beta^2(x_1),\cdots,\widehat{x_{m,n,i}},\cdots,\beta^2(x_{k+2}))\nonumber\\
&&+\sum_{m<i<n}(-1)^{m+n+i+1}\phi^{k+2+s}\rho(x_i)\phi^{-k-3-s}\eta([\beta(x_i),\beta(x_j)],\beta^2(x_1),\cdots,\widehat{x_{m,i,n}},\cdots,\beta^2(x_{k+2}))\nonumber\\
&&+\sum_{i<m<n}(-1)^{m+n+i}\beta^{k+2+s}\rho(x_i)\beta^{-k-3-s}\eta([\beta(x_i),\beta(x_j)],\beta^2(x_1),\cdots,\widehat{x_{i,m,n}},\cdots,\beta^2(x_{k+2}))\nonumber\\
&&+\sum_{i<j}(-1)^{i+j}\phi^{k-1+s}\rho([\beta^2(x_i),\beta^2(x_j)])\phi\phi^{-k-1-s}\eta(\beta^2(x_1),\cdots,\widehat{x_{i,j}},\cdots,\beta^2(x_{k+2}))\label{eq7}\\
&&+\sum_{p<i<j}(-1)^{p+i+j+1}\phi^{k+2+s}\rho(x_p)\phi^{-k-3-s}\eta([\beta(x_i),\beta(x_j)],\beta^2(x_1),\cdots,\widehat{x_{p,i,j}},\cdots,\beta^2(x_{k+2}))\nonumber\\
&&+\sum_{i<p<j}(-1)^{p+i+j}\phi^{k+2+s}\rho(x_p)\phi^{-k-3-s}\eta([\beta(x_i),\beta(x_j)],\beta^2(x_1),\cdots,\widehat{x_{i,p,j}},\cdots,\beta^2(x_{k+2}))\nonumber\\
&&+\sum_{i<j<p}(-1)^{p+i+j+1}\phi^{k+2+s}\rho(x_p)\phi^{-k-3-s}\eta([\beta(x_i),\beta(x_j)],\beta^2(x_1),\cdots,\widehat{x_{i,j,p}},\cdots,\beta^2(x_{k+2}))\nonumber.
\end{eqnarray}
By $\rho([x,y])\phi=\rho(\beta(x))\rho(y)-\rho(\beta(y))\rho(x)$, we have
$$(\ref{eq5})+(\ref{eq6})+(\ref{eq7})=0.$$
About above equations, sum of the rest six equations is zero. So, we proof that $d^s\circ d^s=0$.\qed

\section{Skew-Hom-Lie algebras in semi-Euclidean 4-spaces}
Let$R^4$ be a 4-dimensional vector space.For any two vectors
 $x = (x_1,x_2,x_3,x_4)^T$ and $y = (y_1,y_2,y_3,y_4)^T$ in $R^4$, their
 pseudoscalar product is defined by
 $$<x,y> =x^T\begin{pmatrix}-1&0&0&0\\0&-1&0&0\\0&0&1&0\\0&0&0&1\end{pmatrix}y= -x_1 y_1-x_2 y_2+x_3 y_3+x_4 y_4. $$
 The space $(R^4,<\cdot,\cdot>) $is called semi-Euclidean 4-space with
 index 2 and denoted by $R^4_2$. A non zero vector $x\in R^4_2$ is called spacelike, null,
 or timelike if$<x,x>>0,\quad <x,x>=0$, or $<x,x><0$, respectively. The null space of semi-Euclidean 4-spaces is $V_0=\{x\in R^4_2|<x,x>=0\}$. The null space $V_0$ is not a vector space.

$(\frkg\frkl(V),[\cdot,\cdot]_\alpha,Ad_\alpha)$ is the skew-Hom-Lie algebra in Section 2. When $V$ is the vector space $R^2$,
let $$e_{11}=\begin{pmatrix}1&0\\0&0\end{pmatrix},e_{12}=\begin{pmatrix}0&1\\0&0\end{pmatrix},e_{21}=\begin{pmatrix}0&0\\1&0\end{pmatrix},
e_{22}=\begin{pmatrix}0&0\\0&1\end{pmatrix},$$
$$Ad_\alpha(e_{11},e_{12},e_{21},e_{22})=(e_{11},e_{12},e_{21},e_{22})P,$$for $B\in\frkg\frkl(R^2)$,
we have: $Ad_\alpha\circ Ad_\alpha(B)=B$. Thus, we have:$$P^2=\rm{id}.$$ For the map
$Ad_\alpha$,\quad we named $P$ is the corresponding map.

Let $\alpha=\begin{pmatrix}-\theta & \sqrt{1+\theta^2}\\
-\sqrt{1+\theta^2} & \theta\end{pmatrix}$, $\theta\in \mathbb{R}$,
then $\alpha^2=-\rm{id}$ and
\begin{equation*}
\alpha\circ\alpha^T=\begin{pmatrix}
1+2\theta^2 & 2\theta\sqrt{1+\theta^2}\\
2\theta\sqrt{1+\theta^2} & 1+2\theta^2
\end{pmatrix},
\end{equation*}
when $\theta\neq 0$, $\alpha$ is not an orthogonal matrix. The corresponding map $P$ is
$$\begin{pmatrix}
\theta^2 & \theta \sqrt{1+\theta^2} &-\theta \sqrt{1+\theta^2} & -1-\theta^2\\
-\theta \sqrt{1+\theta^2} & -\theta^2 & 1+\theta^2 & \theta \sqrt{1+\theta^2}\\
\theta \sqrt{1+\theta^2} & 1+\theta^2 & -\theta^2 & -\theta \sqrt{1+\theta^2}\\
-1-\theta^2 & -\theta \sqrt{1+\theta^2} & \theta \sqrt{1+\theta^2} & \theta^2
\end{pmatrix}.$$

And when $\theta\neq 0$, $P$ is not an orthogonal matrix.  Let $r=(-\theta,\sqrt{1+\theta^2},-\sqrt{1+\theta^2},\theta)^T$, by straightforward computations,we have $Pr=-r$.
For $x=(x_1,x_2,x_3,x_4)^T,y=(y_1,y_2,y_3,y_4)^T$,we define $[\cdot,\cdot]_\theta$by
$$[x,y]_\theta=Px\wedge r\wedge y-Py\wedge r\wedge x.$$
where $$Px\wedge r\wedge y=\begin{vmatrix}e_1&-e_2&e_3&e_4\\
                &&(Px)^T&\\
                &&r^T&\\
                &&y^T&
\end{vmatrix},$$ and $\{e_1,e_2,e_3,e_4\}$ is the canonical basis of $R^4$.
By straightforward computations,
\begin{eqnarray*}
&&Px\wedge r\wedge y\\
&=&(-\sqrt{1+\theta^2}(x_1y_3+x_1y_2+x_3y_4+x_2y_4)+\theta(y_2x_3-x_2y_3),\\
&&\theta(x_1y_3+x_3y_1+x_3y_4+x_4y_3)+\sqrt{1+\theta^2}(x_4y_4-x_1y_1),\\
&&\theta(x_1y_2+x_2y_1+x_2y_4+x_4y_2)+\sqrt{1+\theta^2}(x_1y_1-x_4y_4),\\
&&\sqrt{1+\theta^2}(x_2y_1+x_3y_1+x_4y_2+x_4y_3)+\theta(x_3y_2-x_2y_3))^T
\end{eqnarray*}
then $[x,y]_\theta=(a,0,0,a)^T$,where $$a=-\sqrt{1+\theta^2}[(x_1-x_4)(y_2+y_3)-(x_2+x_3)(y_1-y_4)].$$
Because of $P^2=\rm{id}$,
$$[Px,Py]_\theta=x\wedge r\wedge Py-y\wedge r\wedge Px=Px\wedge r\wedge y-Py\wedge r\wedge x=[x,y]_\theta,$$then we have:
$$P([x,y]_\theta)=P(a,0,0,a)^T=-(a,0,0,a)^T=-[x,y]_\theta=-[Px,Py]_\theta.$$

For $z\in R^4$,
\begin{eqnarray*}
&&[[x,y]_\theta,Pz]_\theta\\
&=&P([x,y]_\theta)\wedge r\wedge Pz-PPz\wedge r\wedge [x,y]_\theta\\
&=&\begin{pmatrix}-a\\0\\0\\-a
\end{pmatrix}\wedge r\wedge Pz-z\wedge r\wedge\begin{pmatrix}a\\0\\0\\a
\end{pmatrix}\\
&=&(Pz-z)\wedge r\wedge\begin{pmatrix}a\\0\\0\\a
\end{pmatrix}\\
&=&\begin{vmatrix}e_1&-e_2&e_3&e_4\\
b_1&b_2&b_3&b_4\\
-\theta&\sqrt{1+\theta^2}&-\sqrt{1+\theta^2}&\theta\\
a&0&0&a
\end{vmatrix}\\
&=&\begin{pmatrix}-a\sqrt{1+\theta^2}(b_2+b_3)\\
a(2\theta b_3+\sqrt{1+\theta^2}(b_4-b_1))\\
a(2\theta b_2+\sqrt{1+\theta^2}(b_1-b_4))\\
a\sqrt{1+\theta^2}(b_2+b_3)
\end{pmatrix}\\
&=&(0,0,0,0)^T,
\end{eqnarray*}
where
$$ b_1=((\theta^2-1)z_1+\theta \sqrt{1+\theta^2}z_2)-(\theta \sqrt{1+\theta^2}z_3+(1+\theta^2)z_4),$$
$$ b_2=-(\theta \sqrt{1+\theta^2}z_1+(\theta^2+1)z_2)+((1+\theta^2)z_3+\theta \sqrt{1+\theta^2}z_4),$$
$$ b_3=(\theta \sqrt{1+\theta^2}z_1+(1+\theta^2)z_2)-((\theta^2+1)z_3+\theta \sqrt{1+\theta^2}z_4),$$
$$ b_4=-((1+\theta^2)z_1+\theta \sqrt{1+\theta^2}z_2)+(\theta \sqrt{1+\theta^2}z_3+(\theta^2-1)z_4),$$
so the Hom-Jacobi identity is hold. Thus, we have the following:

\begin{pro}
With the above notations, $(R^4_2,[\cdot,\cdot]_\theta,P)$ is a skew-Hom-Lie algebra.
\end{pro}

\begin{thm}
The null space of semi-Euclidean 4-spaces is $V_0=\{x\in R^4_2|<x,x>=0\}$, let $$V^*=\{x=(x_1,x_2,x_3,x_4)^T\in V_0|x_1x_2=x_3x_4\},$$ there exist a skew-Hom-Lie algebra $(R^4_2,[\cdot,\cdot]_\theta,P)$,and such that:
\begin{itemize}
  \item[\rm(1)] $r=(-\theta,\sqrt{1+\theta^2},-\sqrt{1+\theta^2},\theta)^T\in V^*$,and $Pr\in V^*$.
   \item[\rm(2)] for all $x,y\in R^4_2, \quad [x,y]_\theta\in V^*$.
    \item[\rm(3)] for all $z\in V^*, \quad Pz\in V^*$.
\end{itemize}
\end{thm}
\pf
We just need to prove (3), for $z=(z_1,z_2,z_3,z_4)^T\in V^*$,
\begin{eqnarray*}
(Pz)^T\begin{pmatrix}-1&0&0&0\\
0&-1&0&0\\
0&0&1&0\\
0&0&0&1
\end{pmatrix}Pz&=&
z^TP^T\begin{pmatrix}-1&0&0&0\\
0&-1&0&0\\
0&0&1&0\\
0&0&0&1
\end{pmatrix}Pz\\
&=&z^T\begin{pmatrix}1+2\theta^2&2\theta \sqrt{1+\theta^2}&0&0\\
2\theta \sqrt{1+\theta^2}&1+2\theta^2&0&0\\
0&0&-1-2\theta^2&-2\theta \sqrt{1+\theta^2}\\
0&0&-2\theta \sqrt{1+\theta^2}&-1-2\theta^2
\end{pmatrix}z\\
&=&(1+2\theta^2)(z_1^2+z_2^2-z_3^2-z_4^2)+4\theta \sqrt{1+\theta^2}(z_1z_2-z_3z_4)\\
&=&0.
\end{eqnarray*}
So, $Pz\in V_0$, and by straightforward computations,
\begin{eqnarray*}
Pz&=&\begin{pmatrix}
\theta^2 & \theta \sqrt{1+\theta^2} &-\theta \sqrt{1+\theta^2} & -1-\theta^2\\
-\theta \sqrt{1+\theta^2} & -\theta^2 & 1+\theta^2 & \theta \sqrt{1+\theta^2}\\
\theta \sqrt{1+\theta^2} & 1+\theta^2 & -\theta^2 & -\theta \sqrt{1+\theta^2}\\
-1-\theta^2 & -\theta \sqrt{1+\theta^2} & \theta \sqrt{1+\theta^2} & \theta^2
\end{pmatrix}\begin{pmatrix}z_1\\z_2\\z_3\\z_4
\end{pmatrix}\\
&=&\begin{pmatrix}(\theta^2z_1+\theta \sqrt{1+\theta^2}z_2)-(\theta \sqrt{1+\theta^2}z_3+(1+\theta^2)z_4)\\
-(\theta \sqrt{1+\theta^2}z_1+\theta^2z_2)+((1+\theta^2)z_3+\theta \sqrt{1+\theta^2}z_4)\\
(\theta \sqrt{1+\theta^2}z_1+(1+\theta^2)z_2)-(\theta^2z_3+\theta \sqrt{1+\theta^2}z_4)\\
-((1+\theta^2)z_1+\theta \sqrt{1+\theta^2}z_2)+(\theta \sqrt{1+\theta^2}z_3+\theta^2z_4)
\end{pmatrix}.
\end{eqnarray*}
By $z_1^2+z_2^2=z_3^2+z_4^2$ and $z_1z_2=z_3z_4$, we have: $Pz\in V^*$.
\qed

\end{document}